**Synergistic interactions between DNA and actin trigger emergent viscoelastic behavior**

Robert Fitzpatrick[a], Davide Michieletto[b], Karthik R. Peddireddy[a], Cole Hauer[a], Carl Kyrillos[a], Bekele J. Gurmessa[a], and Rae M. Robertson-Anderson[a]

[a]Department of Physics and Biophysics, University of San Diego, San Diego, CA, USA
[b]School of Physics and Astronomy, University of Edinburgh, Edinburgh, UK

**Abstract:** Composites of flexible and rigid polymers are ubiquitous in biology and industry alike, yet the physical principles determining their mechanical properties are far from understood. Here, we couple force spectroscopy with large-scale Brownian Dynamics simulations to elucidate the unique viscoelastic properties of custom-engineered blends of entangled flexible DNA molecules and semiflexible actin filaments. We show that composites exhibit enhanced stress-stiffening and prolonged mechano-memory compared to systems of actin or DNA alone, and that these nonlinear features display a surprising non-monotonic dependence on the fraction of actin in the composite. Simulations reveal that these counterintuitive results arise from synergistic microscale interactions between the two biopolymers. Namely, DNA entropically drives actin filaments to form bundles that stiffen the network but reduce the entanglement density, while a uniform well-connected actin network is required to reinforce the DNA network against yielding and flow. The competition between bundling and connectivity triggers an unexpected stress response that leads equal mass DNA-actin composites to exhibit the most pronounced stress-stiffening and the most long-lived entanglements.

Mixing polymers with distinct structural features and stiffnesses endows composite materials with unique macroscopic properties such as high strength and resilience coupled with low weight and malleability [1-4]. These versatile materials, ranging from carbon nanotube-polymer nanocomposites and liquid crystals to cytoskeleton and mucus, have numerous applications from tissue engineering to high-performance energy-storage [2,5-12]. Compared to single-constituent materials, polymer composites offer a wider dynamic range and increased control over mechanical properties by tuning the relative concentrations and properties of the different species. Importantly, the unique mechanics that emerge in composites often cannot be deduced from those of the corresponding single-component systems [3,13-17]. However, the physical principles that couple structural interactions to mechanics in composites remain elusive.

Over the past two decades, DNA and actin have been extensively studied as model polymer systems [18-22]. While the contour lengths of each biopolymer can be comparable ($L \approx 10$–50 μm), actin is much stiffer than DNA with a persistence length $l_p$ of ~10 μm compared to $l_p \approx 50$ nm for DNA. When sufficiently long, both polymers form entangled networks over similar concentrations ($c \approx 0.1$-2.5 mg/ml), with actin forming nematic domains above 2.5 mg/ml [18]. Despite their wide use as model systems, very few studies have examined composites of actin and DNA, focusing solely on steady-state structure at concentrations above the nematic crossover or under microscale confinement [23-25]. These studies reported large-scale phase separation such that DNA and actin polymers were rarely interacting. Co-entangled systems of DNA and actin have yet to be investigated.

Here, we directly address these open problems by using optical tweezers microrheology and Brownian Dynamics (BD) simulations to characterize the microscale structure, nonlinear mechanical response, and relaxation dynamics of custom-engineered composites of entangled DNA and actin. We reveal a surprising non-monotonic dependence of stiffening and mechano-memory on composite composition. BD simulations show that these emergent properties arise from a competition between DNA-driven actin bundling and actin network connectivity to scaffold DNA.

The dynamics of entangled polymers can often be described by reptation theory [26,27] which models each polymer as being confined to a tube of diameter $a$ formed by the surrounding polymers, restricting diffusion to a direction parallel to the polymer contour. This confinement arises at times longer than the entanglement time $\tau_e$ (i.e the time needed for polymer segments to reach the tube edge). To relax induced strain, polymers reptate out of deformed tubes over the disengagement time $\tau_D$. Theoretical predictions for these length and timescales are highly dependent on whether the polymer is considered a flexible random coil ($L>>l_p$) or an extended semiflexible polymer ($L\sim l_p$) (see SM) [27-30].

We have designed entangled DNA-actin composites with varying mass fractions of actin $\Phi_A=c_A/(c_A+c_D)$ and a fixed concentration $c=c_A+c_D=0.8$ mg/ml (Fig 1, SM) [31], judiciously chosen such that $a$ and $\tau_e$ for actin-only and DNA-only systems are nearly identical ($a\approx 0.76$ μm, $\tau_e\approx 0.04$ s) [27-30,32-34]. Polymer lengths were chosen such that the primitive path length (or tube length) of flexible DNA, $L_{0,D}\approx 5$ μm [27,32], is comparable to the extended actin contour length ($L_A\approx 7$ μm) [35]. Thus, as we vary $\Phi_A$ we are only changing the mass fraction of flexible and semiflexible polymers while fixing the other system parameters (see SM).

For microrheology measurements, a microsphere is optically displaced 30 μm through the composite at 20 μm/s while the force the composite exerts on the bead during and after strain is measured (Figs 1, S1) [36-38]. During strain, force curves for all networks exhibit three distinct regimes: an initial steep (elastic) increase until $t_1\approx 0.04$ s; a shallower power-law rise $F\sim x^{\alpha_1}$; and a largely viscous regime with $F\sim x^{\alpha_2}$, where $\alpha_2$ approaches zero (Fig 2A). However, there is a clear distinction between composites ($0<\Phi_A<1$) and actin-only ($\Phi_A=1$) or DNA-only ($\Phi_A=0$) networks. Upon normalization of each curve by its terminal value $F_t$, all composites collapse to a universal curve that exhibits more sustained elasticity than single-component networks, with $\alpha_1\approx 0.46$ and $\alpha_2\approx 0.18$ versus $\alpha_1\approx 0.35$ and $\alpha_2\approx 0$ for single-component systems (Fig 2A,C). To further quantify the time-dependent elasticity or stiffness, we compute the effective differential modulus $K=dF/dx$. As shown (Fig 2B), all composites stress-stiffen ($dK/dx>0$) from an initial value $K_0$ to a maximum value $K_{max}$, followed by stress-softening ($dK/dx<0$) and yielding. However, the degree of stiffening ($K_{max}/K_0$) and the lengthscale over which stiffening occurs, $x_{stiff}=x(K_{max})$, display a non-monotonic dependence on $\Phi_A$ (Fig 2B,D). Composites exhibit increased and prolonged stiffening compared to single-component systems, with a maximum in $K_{max}/K_0$ and $x_{stiff}$ observed in equal mass composites ($\Phi_A=0.5$). While the timescale to yield to the viscous regime, $t_y$ (i.e. $t$ at which $K=K_0/2e$ [36,39]), is close to the first crossover time $t_1$ for all systems, $t_y$ reaches a maximum at $\Phi_A=0.5$ (Fig 2E). Finally, the terminal $K$ value, which quantifies the sustained stiffness, displays the signature non-monotonicity, with $\Phi_A=0.5$ exhibiting the most pronounced terminal elasticity (Fig 2E).

Following strain, force relaxation curves for composites also exhibit three distinct regimes with similar crossover times to those during strain: an initial stalling period with minimal force dissipation until $t_1\approx 0.04$ s, power-law relaxation with a $\Phi_A$-independent scaling exponent $\beta_1\approx 2/3$ until $t_2\approx 0.5$ s, followed by more shallow decay with scaling $\beta_2\approx 1/3$ (Fig 3). Conversely, $\Phi_A=0$ and $\Phi_A=1$ systems undergo fast relaxation (minimal stalling) until $t_1\approx 0.04$ s, followed by a single decay regime with polymer-specific exponents $\beta_{2A}\approx 0.36$ and $\beta_{2D}\approx 0.15$. These emergent properties suggest that synergistic interactions between DNA and actin confer composites with increased mechano-memory and more ordered mechanical response [40-42].

The crossover time $t_1$, mediating the onset of more viscous response and relaxation during and following strain, is remarkably close to the entanglement time $\tau_e \approx 0.04$ s. For $t<\tau_e$, entangled polymers are predicted to relax primarily via bending and stretching modes, whereas for $t>\tau_e$ reptation is the principal mechanism. The force-stalling phenomenon, coupled with increased stiffening and reduced yielding during strain, all of which occur at $t<\tau_e$, suggest that bending/stretching is suppressed in composites. The scaling of the second decay phase for composites is similar to that for the actin network, indicating that long-time relaxation is dominated by the slower reptation of actin compared to DNA. While the second crossover time $t_2$ is shorter than the predicted $\tau_D$ for DNA, nonlinear strains have been predicted to dilate entanglement tubes and concomitantly reduce $\tau_D$ [36,37,43-45]. Likewise, during strain composites transition to a primarily viscous regime at $\sim t_2$, (Fig 2A,C), as much of the stress has been relieved via DNA reptation.

To determine the extent to which our results are distinct to the nonlinear regime, we compute the linear elastic modulus $G'(\omega)$ by evaluating the thermal fluctuations of the trapped bead (see SM, [14,46-50]). All networks exhibit a rise in $G'(\omega)$ over a range of $\sim$13–150 rad/s, comparable to the timescales $t_2$ and $t_1$; and $G'(\omega)$ for $\Phi_A$=0.25 and $\Phi_A$=0.75 are similar to that of DNA-only and actin-only networks, respectively (Fig S2). However, $G'(\omega)$ for $\Phi_A$=0.5 exhibits a larger increase with $\omega$, which occurs at higher $\omega$ (shorter $t$) than the other networks. Further, at high $\omega$, $G'(\omega)$ is greatest for $\Phi_A$=0.5 indicating that this system has the most pronounced elastic response to fast strains, in line with our nonlinear regime results (Fig S2).

To shed light on the structural interactions responsible for the emergent stiffening and mechano-memory, we perform large-scale BD simulations (see SM [51,52]). As shown (Figs 1, S3), DNA and actin form networks that span the composite. However, zooming in on simulation snapshots shows that $\Phi_A$=1 networks are formed entirely from entanglements between individual filaments, whereas actin in composites form multi-filament bundles, resulting in less dense networks of bundles (Fig 4A).

To quantify the spatial organization of actin and DNA, we compute the radial distribution function $g_{a-b}(r)=<\delta(|r^a_i-r^b_j|-r)>/g_0$, where $r^a_i$ denotes the position of the $i^{th}$ bead belonging to species $a$ and $g_0=4\rho\pi r^2 dr$ is the expected distribution in uniform systems. Comparing $g_{a-b}$ for actin-actin ($g_{A-A}$), actin-DNA ($g_{A-D}$) and DNA-DNA ($g_{D-D}$), reveals that actin self-associates in the presence of DNA, displayed as peaks in $g_{A-A}$ curves at small $r$ (Figs 4B, S4). These peaks are non-existent in the other distributions, showing that individual DNA polymers remain uniformly distributed, and DNA and actin are well-mixed among each other. We also compute the nematic correlation function $\Pi_{a-b}(r)$ (SM, [25,53,54]), which displays very similar dependence on $\Phi_A$ and $r$ as $g_{A-A}(r)$, demonstrating that actin self-association is nematic bundling rather than randomly-oriented clustering (Fig 4C).

To quantify the lengthscales of actin bundling we compute: (i) the distance $r$ at which $g_{A-A}$ achieves a maximum, $r_a(\Phi_A)$, quantifying spacing between filaments in a bundle; and (ii) the decay distance of $\Pi_{A-A}(r)$, $r_b(\Phi_A)$, quantifying bundle thickness (Table S1, Fig 4). We find that bundles become denser and thinner as DNA concentration increases, as both $r_a$ and $r_b$ decrease with decreasing $\Phi_A$. This effect likely arises from the well-known entropic depletion interaction in which DNA drives actin together to maximize its available volume and entropy [55-57]. We also find that $r_b/r_a$ reaches a maximum at $\Phi_A$=0.5, indicating that there are more filaments per bundle compared to composites with less or more DNA. While $\Phi_A$=0.5 bundles are $\sim$30% less dense than for $\Phi_A$=0.25, allowing them to more efficiently form connections with other bundles, they are comprised of $\sim$20% more filaments ($r_b/r_a(0.5)$=1.73 vs $r_b/r_a(0.25)$=1.43), enhancing stiffness. Importantly, this bundling is on a very different scale than previously reported nematic phases in DNA-actin composites [24,25]. In these studies, DNA and actin phase-separated, forming actin-only and DNA-only regions that spanned >50 μm [24]. Here, DNA and actin remain co-entangled and bundles are on the scale of a few filaments ($r_b/r_a<2$). It is noteworthy that such microscale rearrangements and interactions can lead to such distinct changes to viscoelastic properties. The small scale of bundling also limits the ability of fluorescence confocal microscopy methods used in previous studies [24,25] to accurately capture the morphological changes.

These results suggest that our observed non-monotonic trends (Figs 2,3) arise from a competition between increasing bundle stiffness and maintaining actin network connectivity. While more tightly packed bundles produce stiffer actin fibers to reinforce the DNA, the spacing between bundles also increases producing fewer actin network connections with which DNA can entangle. To quantify actin connectivity in composites and its competition with bundling, we first compare $r_a$ values to the theoretical spacing between monomers in a purely uniform system, $l_f=\rho^{-1/3}$ ($\rho$ is monomer density, see SM). When $r_a<l_f$, as for $\Phi_A=0.25$, connections between non-aligned actin filaments (i.e. entanglements) are destroyed in favor of bundling, while for $r_a>l_f$ (as for $\Phi_A=0.75$), connections are largely preserved but bundling is weak. Notably, for $\Phi_A=0.5$, $r_a\approx l_f$, demonstrating a critical point in which bundling and connectivity are optimally balanced. We also evaluate the probability $P_{bond}$ of any two actin filaments to be in contact, using both $r_a$ and $l_f$ as threshold spacings for contact (Fig S6). As shown, $P_{bond}(l_f)$ decreases with increasing $\Phi_A$, demonstrating that the degree of bundling decreases, whereas $P_{bond}(r_a)$ increases, showing that more bundles are connected to one another. Without bundle connectivity, only filaments within the same bundle would contribute to $P_{bond}(r_a)$, whereas if bundles are connected, filaments in different bundles would also contribute, increasing $P_{bond}$. At $\Phi_A=0.5$, $P_{bond}(r_a)\approx P_{bond}(l_f)$, demonstrating once again the unique criticality of this composition.

To further quantify network structure, we evaluate the density fluctuations $\delta\rho/\rho$ in actin networks and the entropy of mixing $\Delta S/S_{max}$ (SM, Figs 4, S6) [58]. We find that both quantities decrease as $\Phi_A$ increases, indicating that at higher $\Phi_A$, actin provides a more uniform, connected scaffold (suppressing spatial density fluctuations). For $\delta\rho/\rho>1$, as for $\Phi_A=0.25$, fluctuations outweigh uniformity as actin bundles form large holes in the scaffold, while for $\delta\rho/\rho<1$ (seen in $\Phi_A=0.75$), uniformity dominates such that bundling cannot appreciably increase network stiffness. Uniquely, for $\Phi_A=0.5$, $\delta\rho/\rho\approx 1$ (Fig 4), corroborating that a careful balance between bundling and uniformity is achieved.

To demonstrate that these synergistic DNA-actin interactions can lead to the experimentally observed emergent viscoelasticity, we quantify the bulk equilibrium stress relaxation $G(t)$ (SM) [59-61]. We find similar scaling exponents to experimental relaxation values for $\Phi_A=1$ ($\alpha_A\approx 1/3$) and $\Phi_A=0$ ($\alpha_D\approx 0.15$); and at short times $G(t)$ for composites ($0<\Phi_A<1$) display $\alpha\approx 2/3$ scaling, quite close to the experimental $\alpha_1$ (Fig 4D, S7). At $t_1\approx 0.04$ s, all networks display a crossover to a slow-decay regime, with nearly all curves displaying similar scaling ($\alpha\approx 1/3$), aligning with our experimental $\alpha_2$. The notable exception is $\Phi_A=0.5$, which exhibits a long-lived entanglement plateau and transitions to terminal behavior at shorter times than the other networks. Our experiments exhibit a similar phenomenon in which the terminal force relaxation value and the high-$\omega$ $G'(\omega)$ plateau are highest for $\Phi_A=0.5$ (Figs 3A, S2). The time at which $G'(\omega)$ transitions to maximal values is also shorter than other networks. These collective results further demonstrate the increased rigidity of this composite compared to other $\Phi_A$ values.

While we find excellent agreement between our experimental and theoretical scaling exponents and crossover time $t_1$, the timescales over which each regime occurs is different. For experimental relaxations, $t_1$ is the crossover from force-stalling to $\alpha_1$ decay, whereas in simulations, it is the crossover from $\alpha_1$ to $\alpha_2$ decay. However, we do not expect $G(t)$ to be identical to experimental relaxation curves as our experiments measure stress relaxation following nonlinear perturbation, whereas $G(t)$ measures the stress dissipation from thermal deformations. Comparing $G(t)$ and $G'(\omega)$ is also not straightforward as experimental $G'(\omega)$ measurements are performed at the microscale while *G(t)* quantifies the bulk response; and previous studies of blends of stiff and flexible polymers have shown that the elastic response is highly dependent on the lengthscale examined [14,17]. Nonetheless, similarities between simulated and experimental curves corroborate that our simulations can capture the dynamics of our experimental system.

In summary, we provide new general evidence for synergistic interactions between stiff and flexible polymers that can result in enhanced stress-stiffening, robust entanglements, and mechano-memory that well exceed that of the corresponding single-component systems. We show that flexible DNA polymers cause semiflexible actin filaments to bundle via entropic forces, which increases the ability of the composite to stiffen in response to strain and resist yielding and relaxation. However, entropic bundling eventually comes at a cost of destroying actin network connectivity required to reinforce the flexible DNA network against flow and allow for long-lived entanglements. Thus, the non-monotonic viscoelastic response observed in experiments and simulations is a direct consequence of the balance between forming tighter bundles and maintaining network connectivity. We expect our collective results to be generally applicable to any composite in which both flexible and stiff polymers are in the entangled regime. If the concentration exceeds that of the nematic crossover for either species then largescale phase separation is expected [24,25]. If the concentration is below that of the entanglement threshold for the (i) stiff or (ii) flexible species then (i) any degree of bundling would destroy connectivity [15,17] and (ii) the flexible network could no longer contribute to bearing mechanical stresses, both critical to the emergent viscoelastic behavior we report. While substantial changes in viscoelasticity in composites are often attributed to largescale phase separation and structural rearrangement, we have shown that molecular-level interactions and entanglements between two distinct polymers can give rise to emergent dynamics. Our collective results reveal new physical phenomena of composite systems, demonstrate the complex interplay between microscale polymer interactions and material properties, and provide a robust biopolymer platform for investigating the physics of polymer composites.

This research was funded by an AFOSR Biomaterials Award (No. FA9550-17-1-0249) and an NSF CAREER Award (No. 1255446) awarded to RMR-A.

7

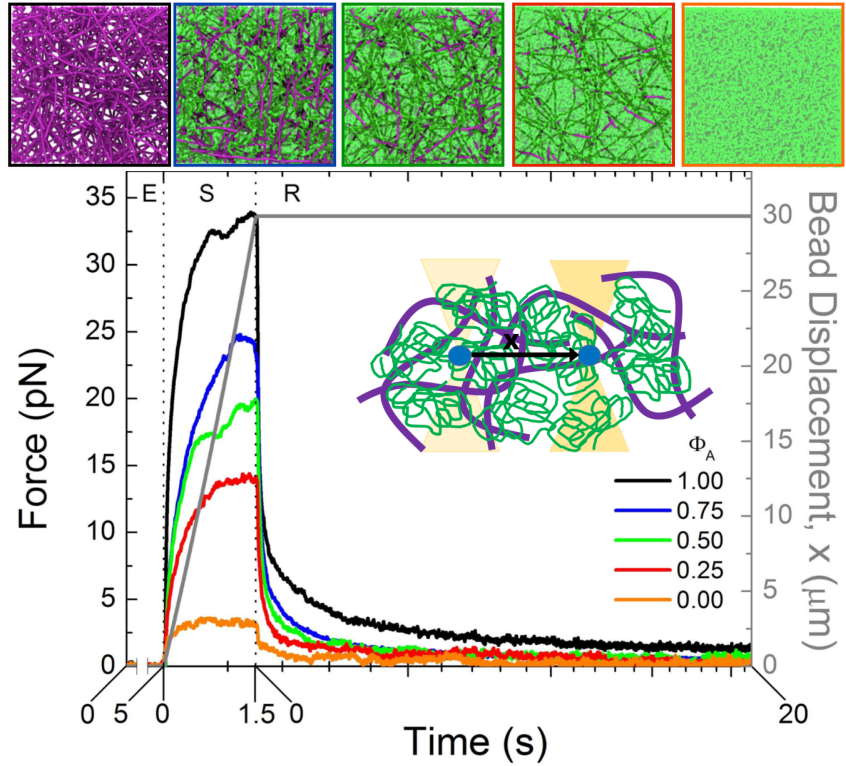

**Figure 1. Optical tweezers microrheology of entangled DNA-actin composites with varying mass fractions of actin, $\Phi_A$.** (Top) Snapshots from BD simulations of entangled composites of actin (magenta) and DNA (green) with varying $\Phi_A$. Each snapshot represents (2.5 μm)$^2$ [(100$\sigma$)$^2$, see SM]. Colors of enclosing boxes signify $\Phi_A$, listed in legend. (Bottom) An optically trapped microsphere (4.5-μm diameter) embedded in the composite is displaced 30 μm (grey) at 20 μm/s. The force is measured before (equilibrium E, 5 s), during (strain S, 1.5 s) and after (relaxation R, 20 s) bead displacement. Each force curve corresponds to a different $\Phi_A$.

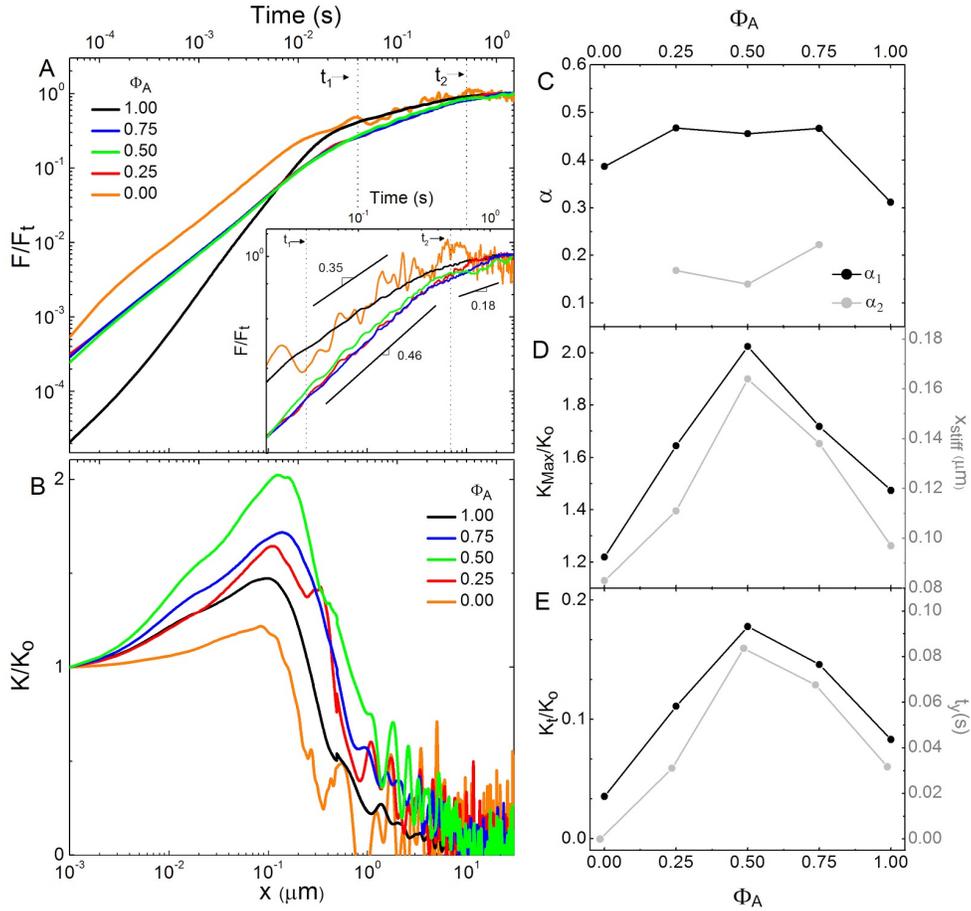

**Figure 2. Equal mass actin-DNA composites display the most pronounced stress-stiffening and resistance to yielding**. (A) Force $F$ as a function of bead displacement $x$ and time $t$, normalized by the terminal value $F_t$, for DNA-actin composites of varying $\Phi_A$. Dashed lines denote times ($t_1$, $t_2$) at which force curves crossover to weaker power-law rise. Inset: Zoom-in of force near the end of strain. Scale bars show average scaling exponents for composites ($\alpha_1 \approx 0.46$, $\alpha_2 \approx 0.18$) and single-component networks ($\alpha_1 \approx 0.35$, $\alpha_2 \approx 0$). (B) Effective differential modulus $K=dF/dx$, normalized by the initial value $K_0$. (C) Dependence of scaling exponents $\alpha_1$ (black) and $\alpha_2$ (grey) on $\Phi_A$. (D) Dependence of stress-stiffening on $\Phi_A$. The maximum differential modulus $K_{max}$, normalized by $K_0$, quantifies the degree to which composites stress-stiffen (black). The bead displacement at which $K_{max}$ is reached, $x_{stiff}$, quantifies the lengthscale over which composites stiffen (grey). (E) Dependence of yielding on $\Phi_A$. The terminal $K$ value, $K_t$, quantifies the amount of stiffness composites retain at the end of the strain (black). The yield time, $t_y$, quantifies the time over which composites lose initial elasticity and yield to a viscous regime (grey).

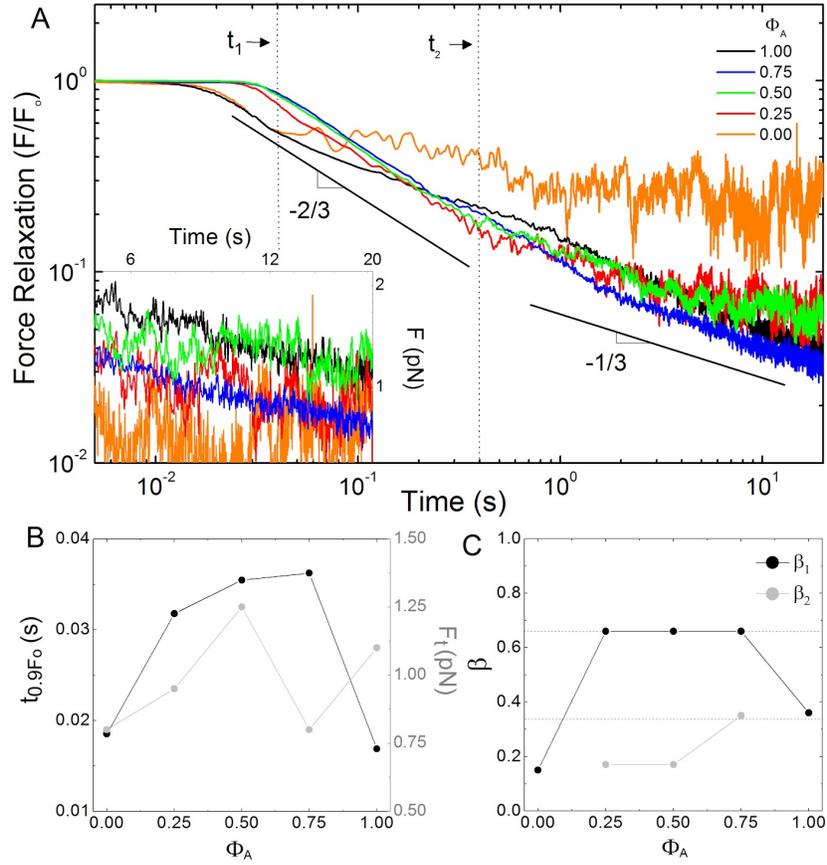

**Figure 3: Composites display universal force-stalling and power-law force relaxation.** (A) Relaxation of force $F$ as a function of time $t$ following strain, normalized by the corresponding force at $t=0$, $F_0$, for networks of varying $\Phi_A$. Black lines indicate power laws, $F \sim t^{-\beta}$, with exponents listed. Composites ($0<\Phi_A<1$) display an initial stalling period until $t_1 \approx 0.04$ s (dashed line), after which power-law relaxation ensues with $\alpha_1 \approx 2/3$. For $t_2 > 0.5$ s (dashed line), relaxation displays a weaker decay with $\alpha_2 \approx 1/3$. Conversely, single-component networks exhibit near immediate relaxation ($t<0.02$ s), with an initial fast decay until $t_1 \approx 0.04$ s followed by single power-law decays. Inset: Un-normalized force at the end of relaxation showing that $\Phi_A=0.5$ composites retain the most force. (B) Stalling time (black), determined as the time at which $F$ drops to $0.9F_0$, and terminal force $F_t$ at the end of relaxation (grey), as a function of $\Phi_A$. (C) Scaling exponents as a function of $\Phi_A$ with dashed lines at 1/3 and 2/3.

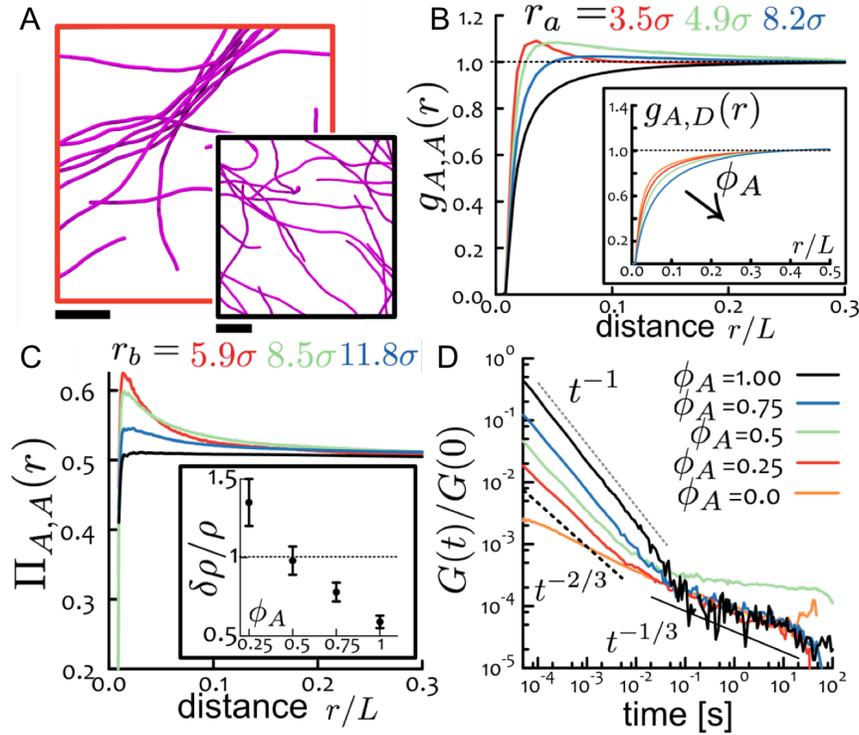

**Figure 4: BD simulations shows that actin bundling causes non-monotonic composite stiffening.** (A) Simulation snapshots showing a trace amount of actin in $\Phi_A$=0.25 (left) and $\Phi_A$=1 (right) composites. Scale bars are 20$\sigma$=500 nm. (B) Radial distribution functions for actin-actin $g_{A\text{-}A}(r)$ and actin-DNA $g_{A\text{-}D}(r)$ (inset), as a function of distance $r$ (normalized by box size $L$) for varying $\Phi_A$. Values of $r_a$ quantify the distance between actin filaments in bundles. (C) Nematic order parameter for actin, $\Pi_{A\text{-}A}(r)$. Inset: Density fluctuations $\delta\rho/\rho$ decrease with increasing $\Phi_A$, reaching ~1 for $\Phi_A$=0.5. Values of $r_b$ quantify the thickness of bundles. (D) Stress relaxation function $G(t)$ showing two distinct power-law decays with crossover at $t_I \approx 0.04$ s. The case $\Phi_A$=0.5 uniquely exhibits a distinct plateau and larger terminal $G(t)$ values.